\documentclass[iop, aplj, tighten]{emulateapj}
\usepackage{amsmath}
\usepackage{amssymb}
\usepackage{apjfonts}
\usepackage{bm}
\usepackage{booktabs}
\usepackage{enumitem}
\usepackage{url}
\usepackage[backref,breaklinks,colorlinks,citecolor=blue]{hyperref}
\usepackage[all]{hypcap}

\received{March, 2018}
\revised{May, 2018}
\accepted{June, 2018}
\shorttitle{EDGES High-Band Results: II.}
\shortauthors{Monsalve et al.}

\begin{document}

\title{Results from EDGES High-Band: II. Constraints on Parameters of Early Galaxies}

\author{
Raul A. Monsalve\altaffilmark{1,2,3},
Bradley Greig\altaffilmark{4,5,6},
Judd D. Bowman\altaffilmark{2},
Andrei Mesinger\altaffilmark{6},
Alan E. E. Rogers\altaffilmark{7},
Thomas J. Mozdzen\altaffilmark{2},
Nicholas S. Kern\altaffilmark{8},
Nivedita Mahesh\altaffilmark{2}
}

\affil{$^1$Center for Astrophysics and Space Astronomy, University of Colorado, Boulder, CO 80309, USA; \href{mailto:Raul.Monsalve@colorado.edu}{Raul.Monsalve@colorado.edu}}
\affil{$^2$School of Earth and Space Exploration, Arizona State University, Tempe, AZ 85287, USA}
\affil{$^3$Facultad de Ingenier\'ia, Universidad Cat\'olica de la Sant\'isima Concepci\'on, Alonso de Ribera 2850, Concepci\'on, Chile}
\affil{$^4$ARC Centre of Excellence for All-Sky Astrophysics in 3 Dimensions (ASTRO 3D), University of Melbourne, VIC 3010, Australia}
\affil{$^5$School of Physics, University of Melbourne, VIC 3010 Australia}
\affil{$^6$Scuola Normale Superiore, Piazza dei Cavalieri 7, I-56126 Pisa, Italy}
\affil{$^7$Haystack Observatory, Massachusetts Institute of Technology, Westford, MA 01886, USA}
\affil{$^8$Department of Astronomy and Radio Astronomy Laboratory, University of California Berkeley, Berkeley, CA 94720, USA}

\begin{abstract}
We use the sky-average spectrum measured by EDGES High-Band ($90-190$ MHz) to constrain parameters of early galaxies independent of the absorption feature at $78$~MHz reported by \citet{bowman2018}. These parameters represent traditional models of cosmic dawn and the epoch of reionization produced with the 21cmFAST simulation code \citep{mesinger2007, mesinger2011}. The parameters considered are: (1) the UV ionizing efficiency ($\zeta$), (2) minimum halo virial temperature hosting efficient star-forming galaxies ($T^{\rm min}_{\rm vir}$), (3) integrated soft-band X-ray luminosity ($L_{\rm X\,<\,2\,keV}/{\rm SFR}$), and (4) minimum X-ray energy escaping the first galaxies ($E_{0}$), corresponding to a typical H${\rm \scriptstyle I}$ column density for attenuation through the interstellar medium. The High-Band spectrum disfavors high values of $T^{\rm min}_{\rm vir}$ and $\zeta$, which correspond to signals with late absorption troughs and sharp reionization transitions. It also disfavors intermediate values of $L_{\rm X\,<\,2\,keV}/{\rm SFR}$, which produce relatively deep and narrow troughs within the band. Specifically, we rule out $39.4<\log_{10}\left(L_{\rm X\,<\,2\,keV}/{\rm SFR}\right)<39.8$ ($95\%$ C.L.). We then combine the EDGES High-Band data with constraints on the electron scattering optical depth from \emph{Planck} and the hydrogen neutral fraction from high-$z$ quasars. This produces a lower degeneracy between $\zeta$ and $T^{\rm min}_{\rm vir}$ than that reported in \citet{greig2017a} using the \emph{Planck} and quasar constraints alone. Our main result in this combined analysis is the estimate $4.5$~$\leq \log_{10}\left(T^{\rm min}_{\rm vir}/\rm K\right)\leq$~$5.7$ ($95\%$ C.L.). We leave for future work the evaluation of $21$~cm models using simultaneously data from EDGES Low- and High-Band.
\end{abstract}


\section{Introduction} 
\label{section_introduction}

The absorption spectral feature detected recently by the EDGES collaboration \citep{bowman2018}, centered at $78$~MHz, is consistent with many expectations for the timing of formation of the first stars \citep{furlanetto2006, pritchard2007, pritchard2010, mesinger2013, greig2017b, cohen2017}. However, the best-fit absorption amplitude, $\sim0.5$~K, is significantly larger than traditional models for the sky-average, or global, 21~cm signal could physically allow. Theoretical models have already been proposed to explain a deep absorption, which either invoke a stronger radiation background \citep[e.g., ][]{feng2018, ewallwice2018} or a colder intergalactic medium (IGM) \citep[e.g., ][]{tashiro2014, munoz2015, barkana2018, barkana2018b, berlin2018, fialkov2018, munoz2018} during cosmic dawn. Other studies have focused on aspects beyond the amplitude of the signal. For instance, \citet{mirocha2018b} explore the apparent conflict between the implied redshift of the absorption feature and high-$z$ projections of ultra-violet galaxy luminosity functions. In parallel to these theoretical works, efforts are underway \citep[e.g.,][]{bernardi2016, price2018, singh2017a, singh2017b} to independently verify the detection, which resulted from the analysis of data gathered with the two EDGES Low-Band instruments. Until the detection is verified, it is important to rigorously evaluate traditional models of high-$z$ galaxies using different measurements.

In this paper we use the EDGES High-Band spectrum presented in \citet{monsalve2017b} to constrain astrophysical parameters of early galaxies from the widely used 21cmFAST simulation code\footnote{\url{http://homepage.sns.it/mesinger/DexM___21cmFAST.html}} \citep{mesinger2007, mesinger2011}. We do not rely on data or results from EDGES Low-Band and do not assume that the absorption feature detected by \citet{bowman2018} corresponds to the cosmological $21$~cm signal.

To derive the astrophysical constraints, we evaluate the consistency between the High-Band spectrum and $10,000$ global $21$ cm signals, each of which is the result of a 21cmFAST simulation. This analysis is conducted in a Bayesian framework that accounts for statistical and systematic uncertainties in the spectrum. It provides one- and two-dimensional probability density functions (PDFs) for the astrophysical parameters after marginalizing over the other 21cmFAST parameters, as well as over parameters of the foreground model. This is a significant improvement over previous works that have reported preliminary astrophysical constraints derived from the evaluation of a limited number of $21$~cm signals \citep{singh2017a, singh2017b}, and works that have only considered phenomenological models for the signal \citep{bernardi2016, monsalve2017b}.

We also compute astrophysical constraints combining the EDGES High-Band spectrum with constraints on: (1) the electron scattering optical depth, $\tau_e$, from \emph{Planck} \citep{planck2016}, and (2) the average fraction of neutral hydrogen, $\bar{x}_{\rm H{\scriptscriptstyle I}}$, from high-$z$ quasars \citep{mcgreer2015, mortlock2011, greig2017c}.

This paper is organized as follows. In Section~\ref{section_observations} we briefly review the EDGES High-Band data. In Section~\ref{section_models} we describe the 21cmFAST astrophysical parameter space explored. In Section~\ref{section_analysis} we describe the analysis and computation of the parameter PDFs. In Section~\ref{section_results} we present and discuss the results. Finally, in Section~\ref{section_conclusion} we summarize this work.

\section{Observations}
\label{section_observations}
EDGES High-Band was a single-antenna, total-power radiometer that measured the sky temperature spectrum in the frequency range $90-190$ MHz, corresponding to $14.8\geq z\geq 6.5$. It observed between 2015 and 2017 from the Murchison Radio-astronomy Observatory (MRO) in Western Australia.

The High-Band spectrum used in this analysis is described in \citet{monsalve2017b}. It was produced by averaging nighttime measurements of the low-foreground sky range $0.26-6.26$~hr local sidereal time, conducted between 2015 September 7 and 2015 October 26. The data were calibrated by removing the time-dependent instrument gain and offset, which were obtained through a combination of (1) a high-accuracy pre-deployment laboratory characterization of the receiver, and (2) field measurements of an internal ambient load and noise source, measured by the receiver in a continuous cycle alternating between the internal sources and the antenna \citep{rogers2012, monsalve2017a}. We removed the effect of small signal reflections between the antenna and receiver input using measurements of their reflection coefficients. We also estimated and removed the effects of antenna and ground loss using measurements, theoretical models, and electromagnetic simulations. Finally, we removed the effect of the antenna beam chromaticity using the electromagnetic model of the beam and a model for the diffuse foreground based on the Haslam $408$ MHz map \citep{haslam1982} and the Guzm\'an $45$ MHz map \citep{guzman2011}. We excised data affected by radio-frequency interference, primarily in the FM band ($87-108$ MHz) and in the channels around $137$~MHz used by the ORBCOMM satellites. The resulting spectrum is dominated by the power-law synchrotron emission of the galaxy. After removal of a best-fit five-term polynomial it has residuals of $17$ mK. For details about the observations, calibration, and data analysis, we point the reader to \citet{rogers2012, monsalve2017a, monsalve2017b}.

\section{Astrophysical Parameters} 
\label{section_models}

To model the $21$~cm signal, we use the publicly available 21cmFAST simulation code \citep{mesinger2007, mesinger2011}, with the model parametrization outlined in \citet{greig2017b}. Although in principle we can access six 21cmFAST astrophysical parameters, in this paper we only explore four parameters whose variations result in perceivable changes in the $21$~cm signal above the measurement uncertainty. These four parameters are briefly described below. For more detailed descriptions we refer the reader to the works cited above.

\begin{enumerate}[wide, labelwidth=!, labelindent=4pt]
\item[$\zeta$:]{The ionizing output of dark matter halos which host galaxies. A constant $\zeta$ corresponds to a constant ionizing luminosity to halo mass for high-$z$ galaxies; even though this is unlikely to be strictly true, it can effectively translate to a halo-mass-averaged ionizing efficiency. It controls the timing of reionization, with increasing values shifting reionization to earlier epochs.}

\item[$T^{\rm min}_{\rm vir}$:] The minimum halo virial temperature hosting efficient star-forming galaxies. Smaller halos are assumed to be unable to support star formation due to feedback and/or inefficient gas cooling. $T^{\rm min}_{\rm vir}$ effectively controls the timing of the epoch of IGM heating and of reionization. Decreasing $T^{\rm min}_{\rm vir}$ produces more abundant, smaller mass galaxies resulting in these cosmic milestones occurring at earlier times and progressing more slowly.

\item[$L_{\rm X\,<\,2\,keV}/{\rm SFR}$:] The X-ray luminosity per star formation rate escaping the galaxy, integrated up to $2$ keV. In this model, X-ray photons escaping into the IGM are responsible for the IGM heating prior to reionization. A lower $L_{\rm X\,<\,2\,keV}/{\rm SFR}$ produces a colder IGM, leading to a delayed and deeper absorption trough in the global 21 cm signal.

\item[$E_{0}$:] The threshold energy for the self-absorption of X-rays by the interstellar medium of the host star-forming galaxy. Decreasing the threshold energy produces more soft X-ray photons which are absorbed closer to the host galaxy leading to inhomogeneous IGM heating. Increasing $E_{0}$ results in inefficient, uniform IGM heating.
\end{enumerate}

\citet{greig2017b} show the sensitivity of the global signal to changes in each of the 21cmFAST parameters. The parameters that least impact the global signal are the maximum horizon for ionizing photons within the ionized regions ($R_{\rm mfp}$) and the spectral index of the X-ray spectral energy distribution ($\alpha_{\rm X}$). Therefore, throughout this work, we adopt the fixed values $R_{\rm mfp}=15.0$~Mpc and $\alpha_{\rm X} = 1.0$. The choice for $R_{\rm mfp}$ is broadly consistent with the sub-grid recombination models of \citet{sobacchi2014}, while the choice for $\alpha_{\rm X}$ roughly corresponds to the spectral index of high-mass X-ray binaries \citep{mineo2012}.

For the remaining four parameters, we take the fiducial ranges outlined in \citet{greig2017b}:
\begin{itemize}
\item $\zeta\in[10,250]$,
\item $T^{\rm min}_{\rm vir}\in[10^{4},10^{6}]$~K,
\item $L_{\rm X\,<\,2\,keV}/{\rm SFR}$ $\in$ $[10^{38},10^{42}]$~erg~s$^{-1}$~M$^{-1}_{\odot}$~yr, 
\item $E_{0}\in[0.1,1.5]$~keV.
\end{itemize}

Using 21cmFAST, we generate $10,000$ unique models of the global $21$ cm signal to span our 4D astrophysical parameter space by randomly drawing from uniform parameter distributions (uniform in $\log_{10}$ scale for $T^{\rm min}_{\rm vir}$ and $L_{\rm X\,<\,2\,keV}/{\rm SFR}$). For each model we simulate a new set of initial conditions, and calculate the global signal from $z=27$ ($\sim50$~MHz) to $z=6$ ($\sim200$~MHz). At each redshift snapshot output by 21cmFAST we use a $300^3$~Mpc$^3$ simulation volume with $200$ voxels per side-length. While alternative sampling approaches such as Latin-Hypercube sampling could be considered, for our purposes, a random distribution is sufficient owing to the relatively small dimensionality (i.e. four).

\section{Analysis}
\label{section_analysis}

To derive constraints on the astrophysical parameters we use a Bayesian approach centered on the computation of the data likelihood for all the astrophysical models. We obtain parameter constraints under two scenarios: (1) using EDGES (High-Band) data only, and (2) applying three observational constraints in addition to the EDGES data. The rest of this section provides details about the analysis.

\subsection{EDGES Data Likelihood}
\label{section_edges_likelihood}

Here we describe the computation of the EDGES data likelihood for the $21$ cm signals associated to each astrophysical model. We start by modeling the uncertainty in the EDGES integrated spectrum as Gaussian and, thus, write the likelihood of the data as

\begin{align}
\mathcal{L}(d|\bm{\theta})=&\frac{1}{\sqrt{(2\pi)^{N_{\nu}}|\Sigma|}}\nonumber\\
&\times\exp{\Big\{-\frac{1}{2}\big[d-m(\bm{\theta})\big]^T\Sigma^{-1}\big[d-m(\bm{\theta})\big]\Big\}},
\label{equation_likelihood_generic}
\end{align}

\noindent where $d$ is our spectrum with $N_{\nu}$ frequency channels, $m$ is the model for the spectrum with parameters $\bm{\theta}$, and $\Sigma$ is the covariance matrix of the data, of size $N_{\nu}\times N_{\nu}$. The model of the spectrum is the sum of the model for the $21$~cm signal and the model for the foreground, which are functions of different parameters:

\begin{equation}
m(\bm{\theta}) = m_{21}(\bm{\theta}_{21}) + m_{\text{fg}}(\bm{\theta}_{\text{fg}}).
\end{equation}

Here, $\bm{\theta}_{21}$ represents the four-element vector of astrophysical parameters associated with the $21$ cm signal, and $\bm{\theta}_{\text{fg}}$ is the vector of foreground parameters.

We model the foreground with the polynomial expression \citep{mozdzen2016, monsalve2017a, monsalve2017b}

\begin{equation}
m_{\text{fg}}(\bm{\theta}_{\text{fg}}) = \sum_{i=0}^{N_{\text{fg}}-1} a_i \nu^{-2.5+i} = A\bm{\theta}_{\text{fg}}.
\label{equation_foreground}
\end{equation}

The $A$ matrix has a size $N_{\nu}\times N_{\text{fg}}$, where the $N_{\text{fg}}$ columns correspond to the $\nu^{-2.5+i}$ basis functions. The first function, with $i=0$, primarily accounts for the contribution to the brightness temperature by Galactic synchrotron radiation. Other terms fit additional spectral structure in the sky or contributions from residual calibration effects. To determine the optimum number of terms in the foreground model over $90-190$~MHz, we compute the weighted residuals rms for a range of $N_{\text{fg}}$ values. We see the rms going through a knee at $N_{\text{fg}}=5$. Specifically, for $N_{\text{fg}}=1,2,3,4$, the rms are $7100$, $758$, $200$, and $57$~mK, respectively. At $N_{\text{fg}}=5$ the rms is $17$~mK, and adding more terms decreases the rms by $<1$~mK. Thus, in this analysis we use $N_{\text{fg}}=5$.

Other models for the diffuse foreground have been proposed. The most popular is the $\log(T) = \sum a_i(\log\nu)^i$ expression, which also tries to capture the temperature, spectral index, and departures of the spectrum from a power law \citep{pritchard2010, harker2012, harker2015, harker2016, voytek2014, bernardi2015, bernardi2016}. Non parametric alternatives have also been suggested \citep{vedantham2014, switzer2014}, some of which rely on training sets from simulated observations using instrument models and all-sky radio maps \citep{tauscher2018}. Unfortunately, the absolute accuracy of the models and maps is not yet high enough to reach agreement between simulations and measurements at the desired level of a few tens of mK or better \citep[i.e.,][]{haslam1982, guzman2011, de_oliveira_costa2008, zheng2017, sathyanarayana2017a}. Although we considered using other models including the linearized physical foreground model introduced in \citet{bowman2018}, whose basis functions approximate the expected contributions from Galactic synchrotron radiation and the ionosphere, we ultimately chose to use Equation~\ref{equation_foreground}. This is a generic expression that with few terms can model our spectrum, which is potentially subject to low-level calibration residuals \citep{mozdzen2016, monsalve2017a}. We note that, although \citet{singh2017a, singh2017b} use `maximally smooth' functions \citep{sathyanarayana2015, sathyanarayana2017b} in laboratory tests and simulations, during the analysis of sky data they also rely on generic polynomials to model their instrument-filtered foreground signal.

We can evaluate Equation~\ref{equation_foreground} analytically at any foreground parameter values. However, the $21$~cm signals in our library are fixed. For this reason we cannot compute Equation~\ref{equation_likelihood_generic} for random values of the full parameter vector (astrophysical + foreground). Instead, to explore the astrophysical parameter space we compute the likelihood for each $21$~cm signal in our library as follows: (i) First, we subtract the $21$~cm signal from the spectrum. (ii) Then, we subtract from (i) the foreground model that best-fits the difference in (i). (iii) Finally, we compute the likelihood for the $21$~cm signal from the residuals in (ii) after marginalizing over the uncertainty in the foreground parameters. This approach represents an alternative to the simultaneous exploration of the full parameter space with techniques such as Markov Chain Monte Carlo (MCMC). We leave the MCMC analysis for future work, as we are currently testing  tools for efficient evaluation of $21$ cm signals for random combinations of astrophysical parameters. 

Now, we provide details of the process outlined in the previous paragraph. Taking advantage of the linearity of the foreground parameters, we compute their maximum likelihood estimator (MLE) as follows:

\begin{equation}
\hat{\bm{\theta}}_{\text{fg}} = \left(A^T\Sigma^{-1}A\right)^{-1}A^T\Sigma^{-1}\left(d - m_{21}(\bm{\theta}_{21})\right).
\end{equation}

This MLE maximizes the likelihood along the foreground parameter axes having conditioned on a selection of astrophysical parameters, $\bm{\theta}_{21}$. The foreground parameters can now be written in terms of the MLE and a deviation term,

\begin{equation}
\bm{\theta}_{\text{fg}} = \hat{\bm{\theta}}_{\text{fg}} + \bm{\delta}_{\text{fg}}.
\end{equation}

Correspondingly, the foreground model can be expanded as follows:

\begin{equation}
m_{\text{fg}}(\bm{\theta}_{\text{fg}}) = m_{\text{fg}}(\hat{\bm{\theta}}_{\text{fg}}) + m_{\text{fg}}(\bm{\delta}_{\text{fg}}) = A\hat{\bm{\theta}}_{\text{fg}} + A\bm{\delta}_{\text{fg}}.
\end{equation}

With this foreground model, whose only free parameters are $\bm{\delta}_{\text{fg}}$, Equation~\ref{equation_likelihood_generic} becomes a function of $\bm{\theta}_{21}$ and $\bm{\delta}_{\text{fg}}$, and can be written as:

\begin{align}
\mathcal{L}(d|\bm{\theta}_{21}, \bm{\delta}_{\text{fg}})& = \frac{1}{\sqrt{(2\pi)^{N_{\nu}}|\Sigma|}}\nonumber \\
\times\exp&{\Big\{-\frac{1}{2}(d_\star- A\bm{\delta}_{\text{fg}})^T\Sigma^{-1}(d_\star-A\bm{\delta}_{\text{fg}})\Big\}},\label{equation_likelihood_deviation}\\
d_{\star} = \;& d - m_{21}(\bm{\theta}_{21}) - m_{\text{fg}}(\hat{\bm{\theta}}_{\text{fg}}). \label{equation_data_star}
\end{align}

We show in Appendix 1 that the marginalization of this likelihood over the foreground parameters, $\bm{\delta}_{\text{fg}}$, gives the following marginal likelihood for the $21$ cm astrophysical model:

\begin{align}
\mathcal{L}(d|\bm{\theta}_{21}) = & \int\mathcal{L}(d|\bm{\theta}_{21}, \bm{\delta}_{\text{fg}}){\text{d}}\bm{\delta}_{\text{fg}} \nonumber \\ 
=&\sqrt{\frac{(2\pi)^{N_{\text{fg}}-N_{\nu}}}{|\Sigma||C^{-1}|}}\exp{\Big\{-\frac{1}{2}d_{\star}^T(\Sigma + V)^{-1}d_{\star}\Big\}},
\label{equation_likelihood_result}
\end{align}

\noindent where $C=(A^T\Sigma^{-1}A)^{-1}$ is the $5\times 5$ covariance matrix of the foreground parameters, $V = (\Sigma_{\text{fg}}^{-1} - \Sigma^{-1})^{-1}$, and $\Sigma_{\text{fg}}=ACA^T$ is the $N_{\nu}\times N_{\nu}$ covariance matrix of the foreground model in the frequency domain. The dependence of $\mathcal{L}$ on the astrophysical parameters is only through $d_{\star}$. $C$ and $V$ have to be calculated only once, as they depend only on (1) the covariance matrix of the data and (2) the foreground model. This enables us to process many $21$ cm signals very efficiently.

The covariance matrix of the data, $\Sigma$, is the sum of the covariance matrix of the noise and a matrix that accounts for systematic uncertainty. The noise covariance matrix is diagonal and describes the channel variance resulting from the effective system temperature and integration time. For reference, the channel standard deviation at $140$ MHz with a channel width of $390.6$ kHz is $6$ mK.

Our systematics covariance matrix represents the variations in the spectral residuals to a five-term foreground model, for different values of the calibration parameters within their uncertainty range. To estimate this matrix we propagate the uncertainty in the calibration parameters to the spectrum in a Monte Carlo (MC) fashion. In each MC realization we calibrate the data by randomly perturbing all the calibration parameters simultaneously in the ranges listed in Table 1 of \citet{monsalve2017b}. Then, we compute the rms residuals across the spectrum to our five-term foreground model. Finally, we identify the $68\%$ upper limit for the MC rms distribution and subtract the minimum rms of the distribution, which occurs for our nominal calibration. From this computation we obtain a $68\%$ limit of $35$ mK. Due to the wide range of perturbation possibilities resulting from all uncertainty sources affecting simultaneously, we categorize the channel-to-channel systematic uncertainty as uncorrelated. Thus, we arrive at the diagonal systematics covariance matrix $\Sigma_{\text{syst}}=(35\;\text{mK})^2\bm{I}_{N_{\nu}}$. We are leaving for future work the estimation of a more refined covariance matrix. See Section~\ref{section_future_work}.

Using a more restrictive foreground model that only accounts for the expected physical foreground contributions could reduce the covariance with the $21$~cm signals. However, it could also lead to (1) higher residuals from the fit to the nominal spectrum if low-level calibration errors are present, and (2) a higher systematic uncertainty estimate, derived as described in the previous paragraph. As an example, fitting the High-Band spectrum with the linearized physically-motivated foreground model of \citet{bowman2018}, Equation~$1$, produces higher residuals rms than Equation~\ref{equation_foreground} with five terms. Thus, using such a model might ultimately result in weaker overall constraints and larger effects from fit residuals. For our model of Equation~\ref{equation_foreground}, these effects are discussed in Sections~\ref{section_edges_only} and \ref{section_combined_constraints}.

\subsection{Additional Observational Constraints}
\label{section_additional_likelihood}

In our second analysis scenario we apply three observational constraints in addition to the EDGES data. They correspond to:
\begin{enumerate}
\item The constraint on the electron scattering optical depth from \emph{Planck}, $\tau_{e} = 0.058 \pm 0.012$ \citep{planck2016}. We model this constraint as Gaussian with center and standard deviation as provided by that estimate.
\item A constraint on $\bar{x}_{\rm H{\scriptscriptstyle I}}$ from the `dark fraction', i.e., the fraction of pixels dark in both Ly$\alpha$ and Ly$\beta$, in the spectra of high-redshift quasars. This constraint is independent from reionization modeling. \citet{mcgreer2015} recently obtained a 1$\sigma$ upper limit of $\bar{x}_{\rm H{\scriptscriptstyle I}} \lesssim 0.06 + 0.05$ at $z=5.9$. We model this limit as a uniform probability distribution over $\bar{x}_{\rm H{\scriptscriptstyle I}} \lesssim 0.06$ and a one-sided Gaussian with $\sigma=0.05$ for $\bar{x}_{\rm H{\scriptscriptstyle I}} > 0.06$.
\item The first\footnote{While preparing this manuscript, \citet{banados2017} reported the detection of a new $z=7.5$ quasar with preliminary IGM constraints consistent with \citet{greig2017c}.} detection ($2\sigma$) of an ongoing reionization, obtained from a Ly$\alpha$ damping wing: the $\bar{x}_{\rm H{\scriptscriptstyle I}}$ PDF from the analysis of the $z=7.1$ quasar ULASJ1120+0641 \citep{mortlock2011, greig2017c} with the `Small H{\scriptsize II}' model\footnote{The reionization morphology of this simulation is characterized by numerous, small cosmic H{\scriptsize II} regions as reionization is driven by haloes with masses of $10^{8}\lesssim M_{h}/M_{\odot} \lesssim 10^9$. The results in this paper are insensitive to this choice of morphology.} of reionization.
\end{enumerate}

In this combined analysis scenario, the likelihood for each astrophysical model is obtained by multiplying the likelihood of the EDGES data for the $21$ cm signal (Section \ref{section_edges_likelihood}), by the likelihood of $\tau_e$, $\bar{x}_{\rm H{\scriptscriptstyle I}}$ at $z=5.9$, and $\bar{x}_{\rm H{\scriptscriptstyle I}}$ at $z=7.1$, for the same model, computed from the constraints in points 1-3 above.

\subsection{Astrophysical PDFs}
\label{section_probabilities}

We interpolate the likelihood computed for our sample of astrophysical parameter values, onto a grid of $40$ equally-spaced bins along each of the four dimensions. The interpolation is performed using an inverse distance weighting (IDW) scheme that uses the simple Shepard's method, where we weigh the 4D distance between the interpolated position and the values originally sampled, $r$, as $w = r^{-12}$. The exponent of $-12$ was selected to minimize numerical artifacts in our interpolated probability space. We tested values $-4$, $-8$, and $-12$, and verified that the choice does not alter the shapes of the PDFs or the marginalized $68\%$ and $95\%$ probability limits.

We then assume uniform priors for the astrophysical parameters (uniform in $\log_{10}$ scale for $T^{\rm min}_{\rm vir}$ and $L_{\rm X\,<\,2\,keV}/{\rm SFR}$) and numerically integrate the gridded 4D likelihood to obtain marginalized 1D and 2D posterior PDFs for each parameter and parameter pair. We smooth these PDFs (both 1D and 2D) with a Gaussian kernel of scale $1.5$ in order to minimize small-scale numerical artifacts that resulted from the interpolation procedure.

\begin{figure*}
\begin{center}
\includegraphics[width=0.8\textwidth]{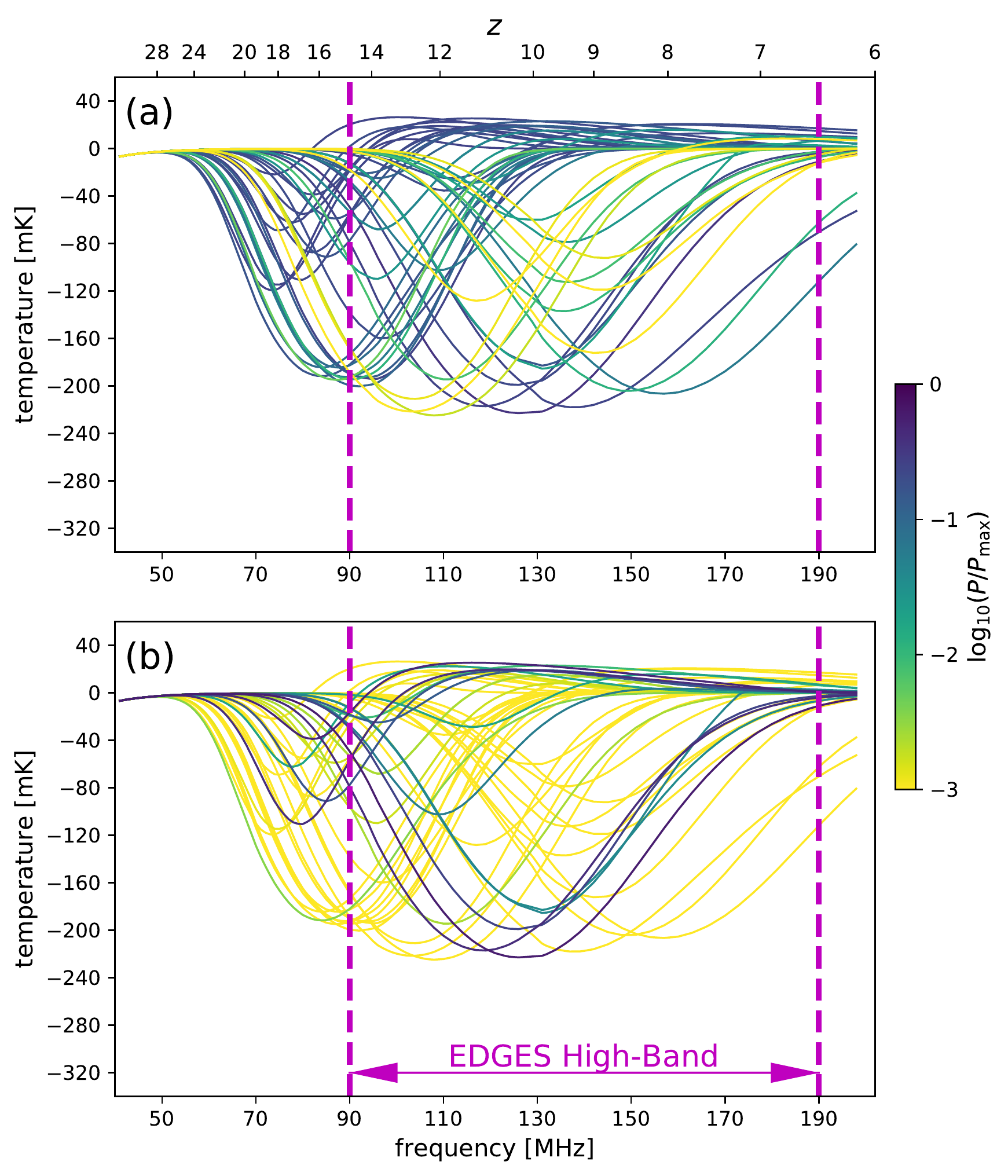}
\caption{Random sample of $50$ global 21 cm signals from our library of $10,000$ signals produced with 21cmFAST. The colors represent the relative probability of each signal, i.e., the original signal probability divided by the highest probability in the library. (a) EDGES data only. The signals with low probability are those with absorption troughs contained mostly within the EDGES band, or troughs that are deep and have sharp features within the band. (b) Combined analysis. The probabilities reflect the significant impact of the constraints on $\tau_e$ and $\bar{x}_{\rm H{\scriptscriptstyle I}}$.  Note that the probability normalization is different between (a) and (b), as the highest probabilities are different in each case. Although in our library there are signals with relative probabilities as low as $\sim 10^{-12}$, to highlight the differences between the signals shown we cap the lower end of the color scale at $10^{-3}$.}
\label{figure_signals}
\end{center}
\end{figure*}

\begin{figure*}
\begin{center}
\includegraphics[width=0.98\textwidth]{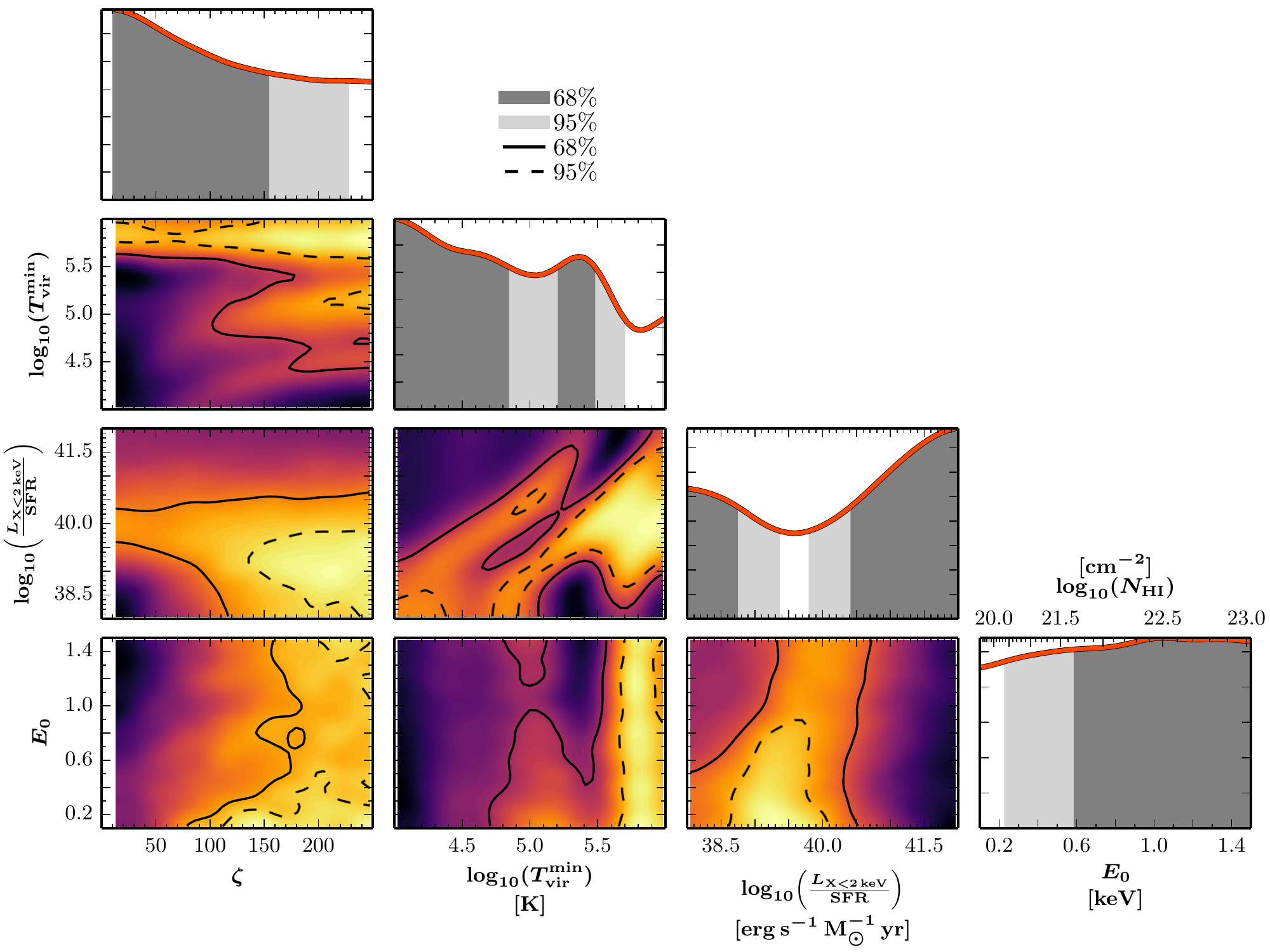}
\caption{Marginalized 1D and 2D posterior PDFs for the 21cmFAST astrophysical parameters explored using EDGES High-Band data. In the 2D PDFs, black (yellow) represents high (low) probability density. We disfavor high values of $T^{\rm min}_{\rm vir}$ and $\zeta$, which mainly regulate the timing of the epoch of heating and reionization. Additionally, we disfavor models with intermediate values of $L_{\rm X\,<\,2\,keV}/{\rm SFR}$ and low values of $E_{0}$. These two parameters affect the timing, width, and depth of the absorption trough in the global signal. We summarize the constraints of this figure in the upper half of Table~\ref{table_limits}. Note that the $68\%$ and $95\%$ probability volumes in the 1D PDFs of $T^{\rm min}_{\rm vir}$ and $L_{\rm X\,<\,2\,keV}/{\rm SFR}$ are contained within two disjoint regions. In Table~\ref{table_limits} we represent these regions as two rows identified by the same letter: `A' for $T^{\rm min}_{\rm vir}$ and `B' for $L_{\rm X\,<\,2\,keV}/{\rm SFR}$.}
\label{figure_edges_data}
\end{center}
\end{figure*}

\begin{figure*}
\begin{center}
\includegraphics[width=0.98\textwidth]{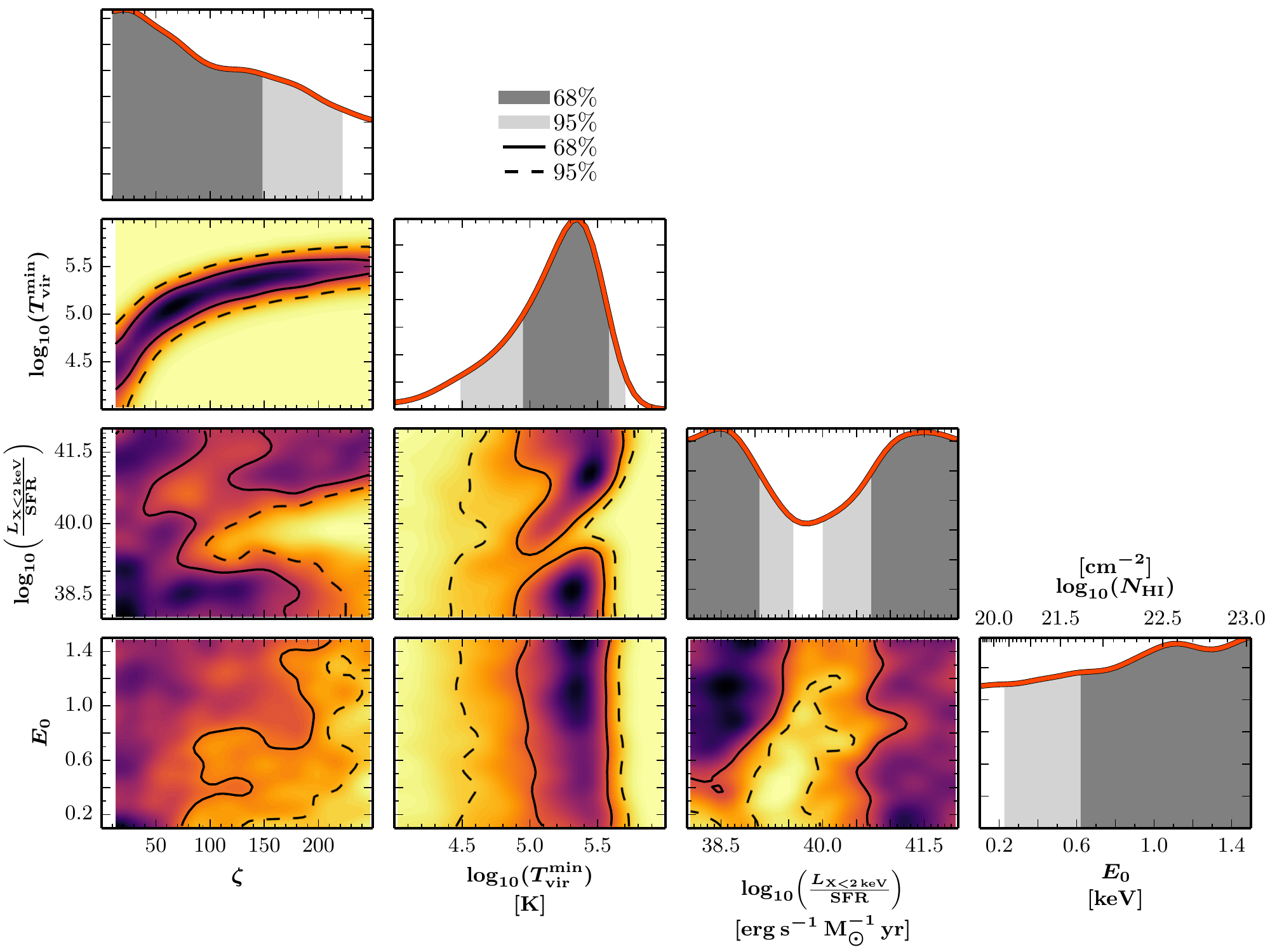}
\caption{PDFs of the astrophysical parameters after incorporating the constraints on $\tau_{e}$ and $\bar{x}_{\rm H{\scriptscriptstyle I}}$ described in Section~\ref{section_additional_likelihood}. In the 2D PDFs, black (yellow) represents high (low) probability density. The most interesting changes relative to the EDGES-only results are those for $\zeta$ and $T^{\rm min}_{\rm vir}$. In particular, the high-probability region in their joint PDF has been significantly constrained. As a reference result, we derive the estimate $4.5$~$\leq \log_{10}\left(T^{\rm min}_{\rm vir}/\rm K\right)\leq$~$5.7$ at $95\%$ confidence. We also see a decrease in the probability of the highest and lowest values of  $L_{\rm X\,<\,2\,keV}/{\rm SFR}$, as well as of low values of $E_0$. The constraints in this figure are summarized in the lower half of Table~\ref{table_limits}. Note that the $68\%$ and $95\%$ probability volumes in the 1D PDF of $L_{\rm X\,<\,2\,keV}/{\rm SFR}$ are contained within two disjoint regions. In Table~\ref{table_limits} we represent these regions as two rows identified with the letter `C'.}
\label{figure_edges_combined}
\end{center}
\end{figure*}

\section{Results and Discussion}
\label{section_results}

The results are summarized in Figures~\ref{figure_signals}$-$\ref{figure_edges_simulation} and in Table~\ref{table_limits}. 

Figure~\ref{figure_signals} shows a random sample of $50$ global signals and their probability from the analyses (a) with EDGES data alone and (b) incorporating the additional constraints on $\tau_e$ and $\bar{x}_{\rm H{\scriptscriptstyle I}}$. We see that the EDGES data disfavor signals with the absorption trough, or relatively sharp parts of the trough, within our band ($90-190$~MHz), consistent with the results in \citet{monsalve2017b}. The $\tau_e$ and $\bar{x}_{\rm H{\scriptscriptstyle I}}$ constraints have a significant effect and reduce the probability of the majority of the signals shown by $\gtrsim3$ orders of magnitude.

We discuss the other figures and the main results next.

\subsection{EDGES Data Only}
\label{section_edges_only}

In Figure~\ref{figure_edges_data} we see that the EDGES High-Band spectrum provides significant discrimination across the parameter space. The most interesting constraints are those for $T^{\rm min}_{\rm vir}$ and $L_{\rm X\,<\,2\,keV}/{\rm SFR}$ (second and third parameters), including their joint PDF, because changes in the value of these parameters have the largest impact on the global signal.

$T^{\rm min}_{\rm vir}$ regulates the minimum mass of halos that produce star-forming galaxies, which strongly affects the timing of the IGM heating and reionization. Lower values of $T^{\rm min}_{\rm vir}$ result in spectral features produced earlier, with a significant fraction of the absorption trough occurring at frequencies below the EDGES High-Band range. Higher values produce troughs to which we have higher sensitivity. This leads to the expectation of lower probabilities as $T^{\rm min}_{\rm vir}$ increases.

$L_{\rm X\,<\,2\,keV}/{\rm SFR}$ controls the IGM heating and, therefore, it has a significant impact on the timing, depth, and width of the absorption trough. For this parameter we expect our highest sensitivity to be below the middle of the range, where the trough is relatively deep and narrow. For high values, the troughs are narrow but shallow, while low values result in late, deep, and wide troughs, as the heating is inefficient and the IGM temperature continues its descending trend before it slowly reacts to the heating.

We see in the $T^{\rm min}_{\rm vir}$ and $L_{\rm X\,<\,2\,keV}/{\rm SFR}$  panels of Figure~\ref{figure_edges_data} that the PDFs are broadly consistent with expectations. The data disfavor high values of $T^{\rm min}_{\rm vir}$ and intermediate values of $L_{\rm X\,<\,2\,keV}/{\rm SFR}$. The high-probability region of low $T^{\rm min}_{\rm vir}$ and high $L_{\rm X\,<\,2\,keV}/{\rm SFR}$, condensed in the upper-left corner of the panel $T^{\rm min}_{\rm vir}-L_{\rm X\,<\,2\,keV}/{\rm SFR}$, corresponds to signals with most of the absorption trough below the EDGES High-Band range. The models in the lower-right corner of the same panel ($\log_{10}\left(T^{\rm min}_{\rm vir}\right)\gtrsim 5.7$ and $\log_{10}\left(L_{\rm X\,<\,2\,keV}/{\rm SFR}\right)\lesssim 39$) have high probabilities because, although they represent ``cold'' EoR scenarios with deep absorption troughs, the troughs fall close to the high-frequency limit of the EDGES band, where we have low sensitivity.

In order to develop a better intuition for the structure in the PDFs of the astrophysical parameters, especially $T^{\rm min}_{\rm vir}$ and $L_{\rm X\,<\,2\,keV}/{\rm SFR}$, we (1) correlated the probability of the $21$~cm signals with the phenomenological parameters of their absorption troughs (center, depth, width), and (2) computed the PDFs for a variety of simulated EDGES High-Band spectra. Specifically, through point (2) we studied the impact of the residuals of the measured spectrum to a five-term foreground model fit. These residuals are shown in Figure 4(b) of \citet{monsalve2017b}, and can be described as ripples across the spectrum with a period of $\approx20$~MHz, an amplitude that decreases with frequency, and a weighted rms over frequency of $17$~mK. In these tests, the simulated spectra consisted of the best-fit foreground model plus noise and ripples that match the real data. In some tests we varied the amplitude of the ripples and in others we added ripples only within specific sub-bands of the spectrum. The simulations also included the systematic uncertainty assigned to the real data. For reference, in Appendix 2, Figure~\ref{figure_edges_simulation}, we show the PDFs resulting from a simulated spectrum that only includes the foreground model and noise. Comparing that figure with Figure~\ref{figure_edges_data} illustrates the impact of the ripples on our results. From our simulations, and refering to Figures~\ref{figure_edges_data} and \ref{figure_edges_simulation}, we derive the following main conclusions: 

\begin{enumerate}
\item The high-probability region of low $T^{\rm min}_{\rm vir}$ and high $L_{\rm X\,<\,2\,keV}/{\rm SFR}$ is not sensitive to the structure in the spectrum above the smooth foreground model.
\item The high-probability blob in the upper-right corner of the panel $T^{\rm min}_{\rm vir}-L_{\rm X\,<\,2\,keV}/{\rm SFR}$ (including its lower-probability tail along the diagonal), is produced by ripples in the measured spectrum below $\sim 110$~MHz. Most signals corresponding to this blob have shallow ($\lesssim 100$ mK) absorption troughs centered around that frequency.
\item Ripples in the range $\sim 110-140$ MHz are responsible for the high-probability intersection of $5.0\lesssim\log_{10}(T^{\rm min}_{\rm vir})\lesssim 5.5$, $\log_{10}\left(L_{\rm X\,<\,2\,keV}/{\rm SFR}\right)\leq 39.5$, and $E_0\gtrsim 0.6$. The corresponding signals have deep ($\gtrsim 200$~mK) but wide ($\rm FWHM \gtrsim 50$~MHz) troughs.
\end{enumerate}

\capstartfalse
\begin{deluxetable}{cccrrrrc}
\tabletypesize{\scriptsize}
\tablewidth{0pt}
\tablecaption{$68\%$ and $95\%$ marginalized limits for the astrophysical parameters. \label{table_limits}}
\tablehead{ & & & \multicolumn{2}{c}{$68\%$} & \multicolumn{2}{c}{$95\%$} &  \\ \cmidrule(lr){4-5}\cmidrule(lr){6-7} \\ \colhead{Parameter} && \colhead{R} & \colhead{Min} & \colhead{Max} & \colhead{Min} & \colhead{Max} & \colhead{Unit}}
\startdata
& & & & & & & \\
\multicolumn{2}{c}{\underline{EDGES Data Only}}  &&&&&&\\
$\zeta$ & \dots          & & $10$ & $154.6$ & $10$ & $228.2$ & \\
$\log_{10}\left(T^{\rm min}_{\rm vir}\right)$        & \dots & \emph{A} & $4$ & $4.9$  & $4$ & $5.7$ & K\\
&& \emph{A} & $5.2$ & $5.5$ & & & K \\
$\log_{10}\left(\frac{L_{\rm X\,<\,2\,keV}}{{\rm SFR}}\right)$         & \dots & \emph{B} & $38$ & $38.8$ & $38$ & $39.4$ & $\frac{\rm erg\;yr}{\rm s \; \rm M_{\sun}}$\\
&& \emph{B} & $40.4$ & $42$ & $39.8$ & $42$ & $\frac{\rm erg\;yr}{\rm s \; \rm M_{\sun}}$\\
$E_0$      & \dots  && $0.58$ & $1.5$ & $0.23$ & $1.5$ & keV\\
&&&&&&\\
\multicolumn{2}{c}{\underline{Combined Analysis}} &&&&&&\\
$\zeta$          & \dots && $10$ & $148.4$ & $10$ & $222.3$ & \\
$\log_{10}\left(T^{\rm min}_{\rm vir}\right)$        & \dots  && $5$ & $5.6$   & $4.5$ & $5.7$ & K\\
$\log_{10}\left(\frac{L_{\rm X\,<\,2\,keV}}{{\rm SFR}}\right)$       & \dots  & \emph{C} & $38$ & $39$ & $38$ & $39.5$  & $\frac{\rm erg\;yr}{\rm s \; \rm M_{\sun}}$\\ %erg $\rm s^{-1}$ $\rm M_{\sun}^{-1}$ yr\\
& & \emph{C} & $40.8$ & $42$ & $40$ & $42$ & $\frac{\rm erg\;yr}{\rm s \rm \;M_{\sun}}$ \\
$E_0$         & \dots && $0.62$ & $1.5$ & $0.23$ & $1.5$ & keV\\
&&&&&&&\\
\enddata
\tablecomments{For some parameters, a given probability volume ($68\%$ or $95\%$) is contained within two disjoint regions due to the shape of their marginalized distribution. Here we represent those disjoint regions as two rows identified by the same letter under the `R' column head.}
\end{deluxetable}

Parameters $\zeta$ and $E_0$ have a lower impact on the global $21$ cm signal but the EDGES data can still discriminate across the value ranges explored. $\zeta$ represents the ionizing efficiency and, thus, has control over the timing and the sharpness of the signal throughout reionization, both while in absorption (reionization happening in parallel to X-ray heating) and emission. As the value of this parameter increases, the reionization transition is sharper and, therefore, higher values are disfavored by the data. $E_0$ represents the low energy limit of the heating X-ray spectrum. It affects the signal in a way related to, but subtler than, $L_{\rm X\,<\,2\,keV}/{\rm SFR}$ \citep{greig2017b}. Lower values of this parameter tend to produce sharper signals and are disfavored by the data. 

The broad high-probability patterns in Figure~\ref{figure_edges_data} are not likely to change significantly with comparable data sets that cover the same frequency range, unless there is a significant reduction in noise level and uncertainty assumed. On the other hand, the high-probability blobs described above for $T^{\rm min}_{\rm vir}$ and $L_{\rm X\,<\,2\,keV}/{\rm SFR}$ could be reduced in the future with a smoother spectrum, assuming that the ripples in the current EDGES High-Band data are due to instrumental artifacts or an incomplete foreground model.

In the upper half of Table~\ref{table_limits} we present the $68\%$ and $95\%$ marginalized limits for the four astrophysical parameters from EDGES data alone. As a reference result, we reject $39.4<\log_{10}\left(L_{\rm X\,<\,2\,keV}/{\rm SFR}\right)<39.8$ at $95\%$ confidence.

\subsection{Combined Constraints}
\label{section_combined_constraints}

Figure~\ref{figure_edges_combined} builds on the findings of \citet{greig2017a}, who derived constraints on the parameters that are most sensitive to the reionization history, $\zeta$, $R_{\text{mfp}}$, and $T^{\rm min}_{\rm vir}$, using constraints on $\tau_e$ from \emph{Planck} and $\bar{x}_{\rm H{\scriptscriptstyle I}}$ from high-$z$ quasars. We point the reader to that work for details and context, in particular their Figures 5 and 8, which are the most relevant for comparison with our results. In our Figure~\ref{figure_edges_combined} we show results after incorporating the same $\tau_e$ and $\bar{x}_{\rm H{\scriptscriptstyle I}}$ constraints to our EDGES analysis. As mentioned in Section~\ref{section_models}, due to the low sensitivity of the global signal to changes in $R_{\text{mfp}}$, we do not include this parameter in our analysis but instead explore $L_{\rm X\,<\,2\,keV}/{\rm SFR}$ and $E_0$, which characterize the IGM X-ray heating.

Out of the four parameters, our interest here focuses on $\zeta$ and $T^{\rm min}_{\rm vir}$. Both regulate the timing of reionization and, thus, suffer from a degeneracy when constrained using just $\tau_e$ and $\bar{x}_{\rm H{\scriptscriptstyle I}}$. As shown in \citet{greig2017a}, these constraints reduce the probability of high $\zeta$ and, even more strongly, the probability of low $T^{\rm min}_{\rm vir}$. We see in Figures~\ref{figure_edges_data} and \ref{figure_edges_combined} that the EDGES measurement contributes to the combined result by disfavoring high values of both, $\zeta$ and $T^{\rm min}_{\rm vir}$. As a result, the joint PDF for these parameters in Figure~\ref{figure_edges_combined} (second row, first column) has a narrower high-probability region than in \citet{greig2017a}. In particular, we estimate $4.5$~$\leq \log_{10}\left(T^{\rm min}_{\rm vir}/\rm K\right)\leq$~$5.7$ at $95\%$ confidence.

The combined analysis disfavors low values of $E_0$ more strongly than EDGES alone. For $L_{\rm X\,<\,2\,keV}/{\rm SFR}$, this analysis still disfavors values in the middle of the range, but the probability of the lowest and, especially, highest values of this parameter is reduced relative to Figure~\ref{figure_edges_data}.

From the simulations described in Section~\ref{section_edges_only} we also identify the effects on Figure~\ref{figure_edges_combined} of the ripples in the spectrum above the foreground model. Specifically, reducing the ripple amplitude at $\sim110$~MHz shifts the high-probability peak at $\log_{10}(L_{\rm X\,<\,2\,keV}/{\rm SFR})\approx41.5$ to slightly higher values (favoring signals with the absorption trough at lower frequencies), while reducing the ripples at $\sim130$~MHz reduces the probability at $\log_{10}(L_{\rm X\,<\,2\,keV}/{\rm SFR})\lesssim39$. All of this also results in lower probability for high values of $\zeta$ and $T^{\rm min}_{\rm vir}$, reducing their covariance.

In the lower half of Table~\ref{table_limits} we present the $68\%$ and $95\%$ marginalized limits from the combined analysis.

\subsection{Probability of the Global $21$~cm Signals}

The combined analysis assigns low probability to a large fraction of global $21$~cm signals. Most of the signals that remain with high probability can be identified with one of the two high-probability regions in the PDF of $L_{\rm X\,<\,2\,keV}/{\rm SFR}$. This results in two main ``bands'' of signals that remain favored. We see hints of this in Figure~\ref{figure_signals}(b), where the majority of the $50$ signals have very low probability (yellow) and the favored signals (green and blue) either have most features at low frequencies, corresponding to high $L_{\rm X\,<\,2\,keV}/{\rm SFR}$, or have troughs that are deep and wide within the EDGES High-Band range, corresponding to low $L_{\rm X\,<\,2\,keV}/{\rm SFR}$.

\begin{figure*}
\begin{center}
\includegraphics[width=0.75\textwidth]{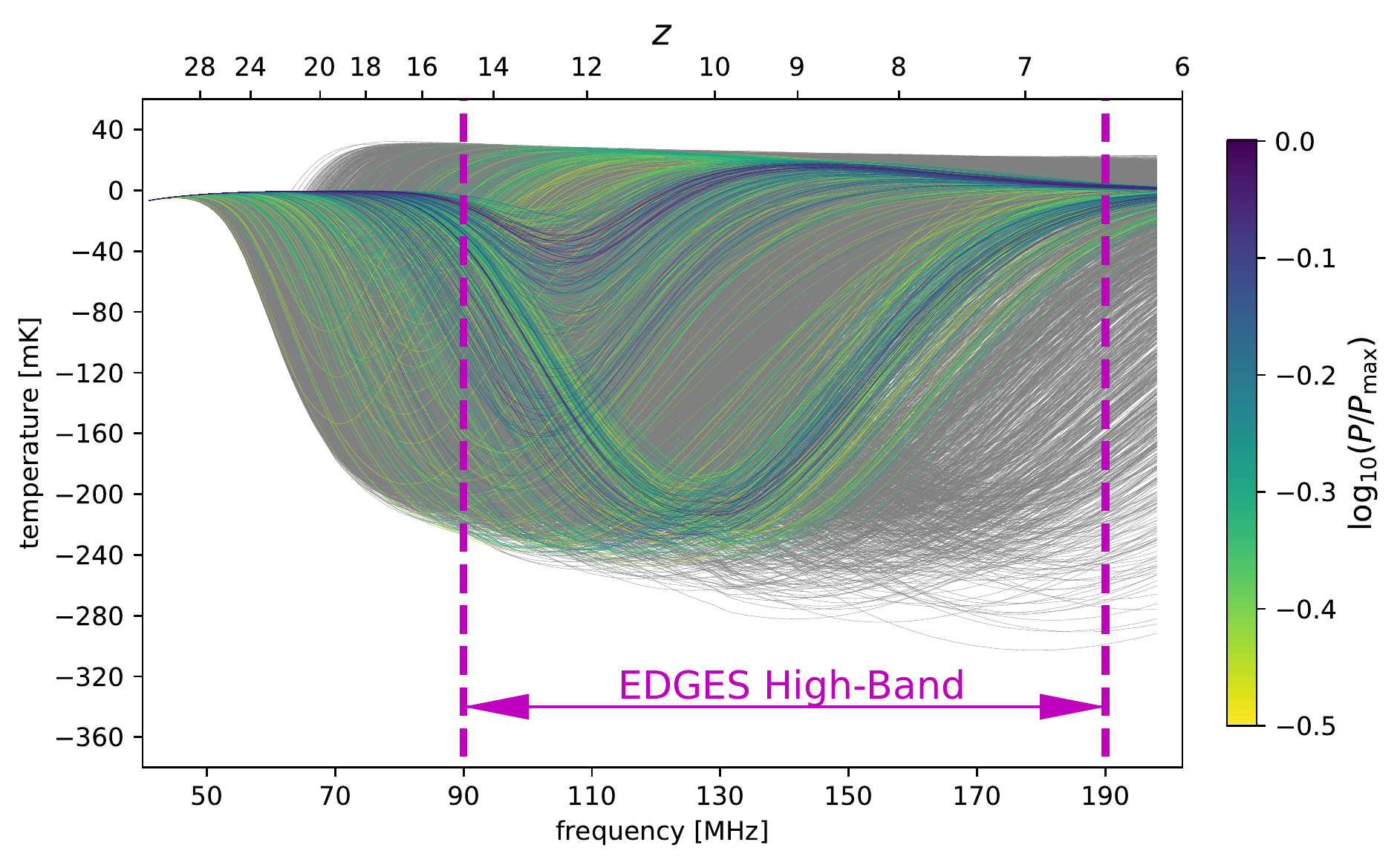}
\caption{In yellow through blue we show the global signals that have the highest $5\%$ probability ($500$ out of $10,000$), from the analysis that combines the EDGES High-Band spectrum and observational constraints on $\tau_e$ and $\bar{x}_{\rm H{\scriptscriptstyle I}}$. The rest of the signals are shown in gray in the background. The two-band pattern observed for the high-probability signals (with absorption troughs centered below or above $\sim110$~MHz) reflects the results for $L_{\rm X\,<\,2\,keV}/{\rm SFR}$ in Figure~\ref{figure_edges_combined}; i.e., the probability is highest for low and high values of this parameter, while middle values are disfavored. Most high-probability signals have brightness temperatures within $\pm25$~mK at $z\approx6$ and can be well fitted by our foreground model over $90-190$~MHz. However, some of those with the absorption trough centered at $\sim105$~MHz and $\sim130$~MHz have a high probability because they match low-level ripples in the spectrum above the foreground model.}
\label{figure_signals_combined_favored}
\end{center}
\end{figure*}

We present a more complete picture of this in Figure~\ref{figure_signals_combined_favored}, where we plot the $10,000$ signals used in this analysis and highlight those that have the highest $5\%$ probability ($500$ signals). These high-probability signals have a brightness temperature within $\pm25$~mK at $z\approx6$ and, although they span a wide range of features, with absorption peaks between $\sim70$ MHz and $\sim130$ MHz, we can identify two prominent high-probability bands: signals with troughs centered below and above $\sim110$~MHz. Most of these signals are smooth over $90-190$~MHz and can be well fitted by our foreground model to within the uncertainties. However, some signals with peaks at $\sim105$ and $\sim130$ MHz are in the top $5\%$ because they match ripples in the spectrum above the foreground model close to those frequencies. As discussed in Sections~\ref{section_edges_only} and \ref{section_combined_constraints}, we identified the effect of this spectral structure through simulations. When this structure is removed in simulated data, the signals with peaks at $\sim105$ and $\sim130$~MHz are replaced in the top $5\%$ by signals with peaks at $\lesssim90$ MHz. Despite this change in some specific signals, the two-band pattern is also present in these simulations of spectrally smooth data, which indicates that this is a robust result from the combined analysis.

In light of the claim by \citet{bowman2018} of an absorption feature centered at $78$~MHz, it is interesting to note that several high-probability signals in Figure~\ref{figure_signals_combined_favored} have troughs at $\sim70-90$ MHz. Regardless of the non-standard depth of the feature detected, having an absorption trough in that range would increase the probability of those signals. However, those signals have $T^{\rm min}_{\rm vir}\sim 10^4$, which is at the corner of our astrophysical parameter space. In our combined analysis that only uses the High-Band spectrum, those values for $T^{\rm min}_{\rm vir}$ are disfavored at $> 95\%$ confidence after marginalization of the posterior over the parameter space (see Figure~\ref{figure_edges_combined} and Table~\ref{table_limits}).

\subsection{Future Work}
\label{section_future_work}

In this paper we have improved on \citet{monsalve2017b} by incorporating the systematic uncertainty in the spectrum into the likelihood computation. However, although the adopted uncertainty level is realistic, its covariance matrix is still generic in that it treats all spectral variations possible within that envelope as equally likely. In order to refine the systematics covariance matrix we plan to conduct a more rigorous characterization of possible spectral perturbation modes of the instrument. This characterization could be joined with analysis techniques that model the instrument passband in terms of orthogonal modes derived from perturbation training sets \citep{burns2017, tauscher2018}. 

We also leave for future work the evaluation of 21cmFAST and other $21$~cm models --- especially those proposed recently to account for the large absorption feature at $78$~MHz --- using simultaneously  data from the EDGES Low- and High-Band instruments, which would provide a total frequency coverage of $50-190$~MHz.

\section{Conclusion}
\label{section_conclusion}

\citet{bowman2018} recently reported the detection of a deep absorption feature in the global spectrum using EDGES Low-Band instruments. In this paper we remain agnostic about this detection and conduct an independent analysis to constrain astrophysical parameters of early galaxies from the 21cmFAST simulation code, using the EDGES High-Band spectrum ($90-190$~MHz). The four parameters explored are $\zeta$, $T^{\rm min}_{\rm vir}$, $L_{\rm X\,<\,2\,keV}/{\rm SFR}$, and $E_0$. We also show constraints for these parameters after incorporating into the analysis observational constraints on $\tau_e$ from \emph{Planck} and $\bar{x}_{\rm H\scriptscriptstyle{I}}$ from high-$z$ quasars.

The astrophysical parameter constraints were computed within a Bayesian framework based on the evaluation of the data likelihood for a wide range of parameter values. The likelihood accounts for estimates of uncertainty in the data due to noise and calibration effects. We obtain the PDF of the astrophysical parameters after assuming uniform priors for these parameters and marginalizing over the five parameters of the foreground model.

We disfavor a wide range of $21$~cm signals with absorption troughs in the EDGES High-Band range, consistent with the expectations from \citet{monsalve2017b}. This translates into significant constraints for the astrophysical parameters. In particular, from EDGES data alone we disfavor high values of $\zeta$ and $T^{\rm min}_{\rm vir}$, and intermediate values of $L_{\rm X\,<\,2\,keV}/{\rm SFR}$. Specifically, we rule out $39.4<\log_{10}\left(L_{\rm X\,<\,2\,keV}/{\rm SFR}\right)<39.8$ at $95\%$ confidence.

Incorporating the constraints on $\tau_e$ and $\bar{x}_{\rm H\scriptscriptstyle{I}}$ into the analysis significantly reduces the high-probability region of the joint PDF for $T^{\rm min}_{\rm vir}$ and $\zeta$. These two parameters control the timing of the epoch of heating and reionization, and suffer from degeneracy. The EDGES High-Band data help reduce this degeneracy relative to previous results that only relied on the $\tau_e$ and $\bar{x}_{\rm H\scriptscriptstyle{I}}$ constraints. From this combined analysis we estimate $4.5$~$\leq \log_{10}\left(T^{\rm min}_{\rm vir}/\rm K\right)\leq$~$5.7$ at $95\%$ confidence.

The signals with highest probabilities from the combined analysis have brightness temperature within $\pm25$~mK at $z\approx6$, and absorption troughs centered below $\sim130$~MHz. Across this signal range we observe two high-probability bands, which have absorption peaks at $\sim70-110$~MHz and $\sim110-130$~MHz, respectively. Although there are several high-probability signals with absorption peaks at $\sim70-90$, in the range of the detected feature \citep{bowman2018}, they correspond to $T^{\rm min}_{\rm vir}\sim 10^4$. After marginalization, this value is disfavored at $>95\%$ confidence by our combined analysis that uses the EDGES High-Band spectrum. We leave for future work the evaluation of traditional and recent models for the global $21$~cm signal using simultaneously data from EDGES Low- and High-Band.

\acknowledgements

We thank Rennan Barkana, Anastasia Fialkov, and Jordan Mirocha for useful discussions. This work was supported by the NSF through research awards for the Experiment to Detect the Global EoR Signature (AST-0905990, AST-1207761, and AST-1609450). R.A.M. was supported by the NASA Solar System Exploration Virtual Institute cooperative agreement 80ARC017M0006, and by the NASA Ames Research Center grant NNX16AF59G. The Centre for All-Sky Astrophysics in 3D (ASTRO 3D) is an Australian Research Council Centre of Excellence, funded by grant CE170100013. This work was supported by the European Research Council (ERC) under the European Union's Horizon 2020 research and innovation programme (grant agreement No 638809 - AIDA - PI: Mesinger). EDGES is located at the Murchison Radio-astronomy Observatory. We acknowledge the Wajarri Yamatji people as the traditional owners of the Observatory site. We thank CSIRO for providing site infrastructure and support.

\emph{Software}: 21cmFAST \citep{mesinger2011}, Astropy \citep{astropy2013}, Ipython (\url{http://dx.doi.org/10.1109/MCSE.2007.53}), Numpy (\url{http://dx.doi.org/10.1109/MCSE.2011.37}), Scipy (\url{https://doi.org/10.5281/zenodo.1036423}), Matplotlib (\url{https://doi.org/10.5281/zenodo.573577}), Healpy \citep{gorski2005}, h5py (\url{https://doi.org/10.5281/zenodo.877338}).

\section*{ORCID \lowercase{i}D\lowercase{s}}
\noindent Raul A. Monsalve \href{https://orcid.org/0000-0002-3287-2327}{https://orcid.org/0000-0002-3287-2327} \\
Bradley Greig \href{https://orcid.org/0000-0002-4085-2094}{https://orcid.org/0000-0002-4085-2094} \\
Judd D. Bowman \href{https://orcid.org/0000-0002-8475-2036}{https://orcid.org/0000-0002-8475-2036} \\
Andrei Mesinger \href{https://orcid.org/0000-0003-3374-1772}{https://orcid.org/0000-0003-3374-1772} \\
Alan E. E. Rogers \href{https://orcid.org/0000-0003-1941-7458}{https://orcid.org/0000-0003-1941-7458} \\
Thomas J. Mozdzen \href{https://orcid.org/0000-0003-4689-4997}{https://orcid.org/0000-0003-4689-4997} \\
Nicholas S. Kern \href{https://orcid.org/0000-0002-8211-1892}{https://orcid.org/0000-0002-8211-1892} \\
Nivedita Mahesh \href{https://orcid.org/0000-0003-2560-8023}{https://orcid.org/0000-0003-2560-8023}

\section*{Appendix 1\\Marginalization of Foreground Parameters}
\label{section_appendix_1}

Here we develop the equation for the marginal likelihood of the $21$ cm signals, presented in Section \ref{section_edges_likelihood}. 

The exponent of the likelihood in Equation \ref{equation_likelihood_deviation} can be written as

\begin{align}
-\frac{1}{2}d_{\star}^T\Sigma^{-1}d_{\star} - \frac{1}{2}\bm{\delta}_{\text{fg}}^TA^T\Sigma^{-1}A\bm{\delta}_{\text{fg}} + d_{\star}^T\Sigma^{-1}A\bm{\delta}_{\text{fg}}.
\label{equation_chi_square_explicit}
\end{align}

The likelihood marginalized over $\bm{\delta}_{\text{fg}}$, therefore, is

\begin{align}
\int\mathcal{L}(d|\bm{\theta}_{21}, \bm{\delta}_{\text{fg}})\text{d}\bm{\delta}_{\text{fg}} = &\frac{1}{\sqrt{(2\pi)^{N_{\nu}}|\Sigma|}}\exp{\Big\{-\frac{1}{2}d_{\star}^T\Sigma^{-1}d_{\star}\Big\}}\nonumber\\&\times\int\exp{\Big\{-\frac{1}{2}\chi_{\dagger}^2\Big\}}\text{d}\bm{\delta}_{\text{fg}}, \label{equation_marginal_likelihood}
\end{align}

\noindent where

\begin{align}
-\frac{1}{2} \chi_{\dagger}^2 =& - \frac{1}{2}\bm{\delta}_{\text{fg}}^TA^T\Sigma^{-1}A\bm{\delta}_{\text{fg}} + d_{\star}^T\Sigma^{-1}A\bm{\delta}_{\text{fg}}. \label{equation_chi_square_dagger}
\end{align}

We can solve this integral (Equation \ref{equation_marginal_likelihood}) by recalling that the integral over a zero-mean multivariate Gaussian distribution plus a linear term is given by \citep{kern2017}

\begin{align}
\int\exp&\left\{-\frac{1}{2}x^TQx + B^Tx\right\}d^nx& \nonumber\\
&= \sqrt{\frac{(2\pi)^n}{|Q|}}\exp\left\{\frac{1}{2}B^TQ^{-1}B\right\}.\label{equation_integral_identity}
\end{align}

We can express $-\frac{1}{2}\chi_{\dagger}^2$ (Equation \ref{equation_chi_square_dagger}) in the form expected on the left-hand side of Equation \ref{equation_integral_identity} by making $x = \bm{\delta}_{\text{fg}}$, $n=N_{\text{fg}}$,  $Q = A^T\Sigma^{-1}A$ and $B = A^T\Sigma^{-1}d_\star$. After applying Equation \ref{equation_integral_identity} with these replacements, we obtain the following marginalized likelihood:

\begin{align}
&\int\mathcal{L}(d|\bm{\theta}_{21}, \bm{\delta}_{\text{fg}})\text{d}\bm{\delta}_{\text{fg}} = \sqrt{\frac{(2\pi)^{N_{\text{fg}}-N_{\nu}}}{|\Sigma||A^T\Sigma^{-1}A|}}\exp{\Big\{-\frac{1}{2}\chi_{\ddagger}^2\Big\}},\label{equation_marginalized_likelihood}\\
&\chi_{\ddagger}^2=d_{\star}^T\Sigma^{-1}d_{\star} -d^T_{\star}\Sigma^{-1}A(A^T\Sigma^{-1}A)^{-1}A^T\Sigma^{-1}d_{\star}. \label{equation_marginalized_chi_square}
\end{align}

In principle, Equations \ref{equation_marginalized_likelihood} and \ref{equation_marginalized_chi_square} are our main result. However, we can rewrite it in a simpler form by making some substitutions. First, we identify $C=(A^T\Sigma^{-1}A)^{-1}$ as the covariance matrix of the foreground parameters, and $\Sigma_{\text{fg}}=ACA^T$ as the foreground covariance projected to the frequency domain. With this, we rewrite Equation \ref{equation_marginalized_chi_square} as:

\begin{align}
\chi_{\ddagger}^2=&d_{\star}^T\left[\Sigma^{-1} - \Sigma^{-1}\Sigma_{\text{fg}}\Sigma^{-1}\right]d_{\star}. \label{equation_marginalized_chi_square_v2}
\end{align}

We can further simplify the expression in brackets by applying the Woodbury matrix identity

\begin{equation}
(\Sigma + V)^{-1} = \Sigma^{-1} - \Sigma^{-1}(\Sigma^{-1} + V^{-1})^{-1}\Sigma^{-1},
\end{equation}

\noindent where $V=(\Sigma^{-1}_{\text{fg}}-\Sigma^{-1})^{-1}$. With this, we arrive at

\begin{equation}
\mathcal{L}(d|\bm{\theta}_{21}) = \sqrt{\frac{(2\pi)^{N_{\text{fg}}-N_{\nu}}}{|\Sigma||C^{-1}|}}\exp{\Big\{-\frac{1}{2}d_{\star}^T(\Sigma + V)^{-1}d_{\star}\Big\}},
\end{equation}

\noindent which is again presented in Equation \ref{equation_likelihood_result}.

\section*{Appendix 2\\PDFs from Simulation}
\label{section_appendix_2}

\begin{figure*}
\begin{center}
\includegraphics[width=0.98\textwidth]{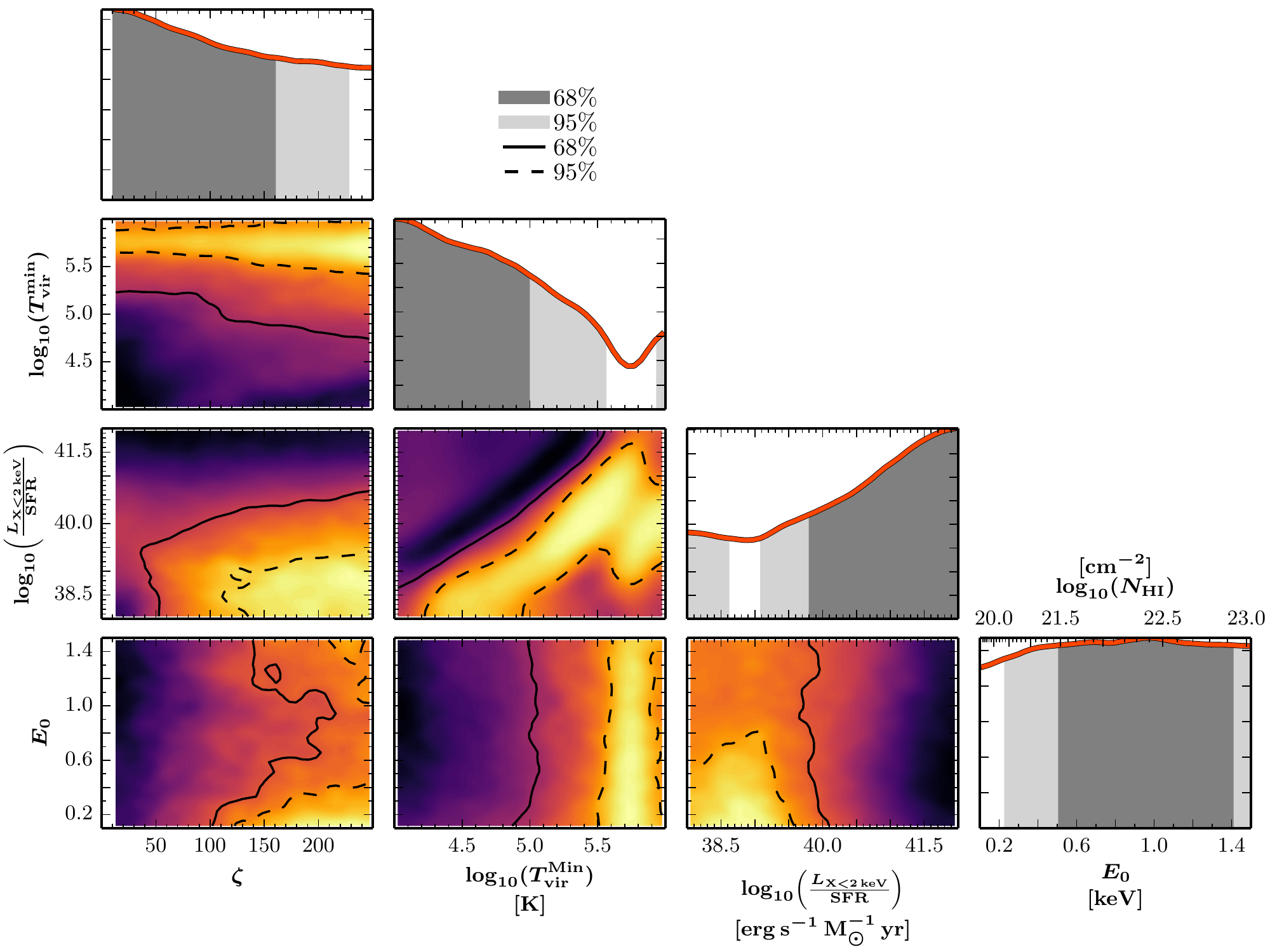}
\caption{Astrophysical parameter PDFs derived from a simulated EDGES High-Band spectrum consisting of the foreground model (Equation~\ref{equation_foreground}) that best fits the real spectrum plus realistic noise. The spectrum does not incorporate ripples or other structure. The computation accounts for the same systematic uncertainty assigned to the real spectrum.}
\label{figure_edges_simulation}
\end{center}
\end{figure*}

In this Appendix we present Figure~\ref{figure_edges_simulation} as reference for the discussion of Section~\ref{section_edges_only}.

{}

\clearpage


\begin{thebibliography}{}

\bibitem[Ba\~nados et al.(2017)]{banados2017} Ba\~nados E., et al. 2017, \href{https://doi.org/10.1038/nature25180}{\nat}, 553, 473

\bibitem[Bernardi et al.(2015)]{bernardi2015} Bernardi, G., McQuinn, M., \& Greenhill, L. J. 2015 \href{https://doi.org/10.1088/0004-637X/799/1/90}{\apj}, 799, 90

\bibitem[Bernardi et al.(2016)]{bernardi2016} Bernardi, G., Zwart, J. T. L., Price, D., et al. 2016, \href{https://doi.org/10.1093/mnras/stw1499}{\mnras}, 461, 3

\bibitem[Barkana(2018)]{barkana2018} Barkana, R. 2018, \href{https://doi.org/10.1038/nature25791}{\nat}, 555, 71

\bibitem[Barkana et al.(2018)]{barkana2018b} Barkana, R., Outmezguine, N. J., Redigolo, D., \& Volansky, T. 2018, \href{https://arxiv.org/abs/1803.03091}{arXiv:1803.03091}

\bibitem[Berlin et al.(2018)]{berlin2018} Berlin, A., Hooper, D., Krnjaic, G., \& McDermott, S. D. 2018, \href{https://arxiv.org/abs/1803.02804}{arXiv:1803.02804}

\bibitem[Bowman et al.(2018)]{bowman2018} Bowman, J. D., Rogers, A. E. E., Monsalve. R. A., Mozdzen, T. J., \& Mahesh, N. 2018, \href{https://doi.org/10.1038/nature25792}{\nat}, 555, 67 

\bibitem[Burns et al.(2017)]{burns2017} Burns, J. O., Bradley, R., Tauscher, K. et al. 2017, \href{https://doi.org/10.3847/1538-4357/aa77f4}{\apj}, 844, 33

\bibitem[Cohen et al.(2017)]{cohen2017} Cohen, A., Fialkov, A., Barkana, R., \& Lotem, M. 2017, \href{https://doi.org/10.1093/mnras/stx2065}{\mnras}, 472, 1915

\bibitem[de Oliveira-Costa et al.(2008)]{de_oliveira_costa2008} de Oliveira-Costa, A., Tegmark, M., Gaensler, B. M., et al. 2008, \href{https://doi.org/10.1111/j.1365-2966.2008.13376.x}{\mnras}, 388, 247

\bibitem[Ewall-Wice et al.(2018)]{ewallwice2018} Ewall-Wice, A., Chang, T., Lazio, T., Dor\'e, O., Seiffert, M., \& Monsalve, R. A. 2018, \href{https://arxiv.org/abs/1803.01815}{arXiv:1803.01815}

\bibitem[Feng \& Holder(2018)]{feng2018} Feng, C., \& Holder, G. 2018, \href{https://arxiv.org/abs/1802.07432}{arXiv:1802.07432}

\bibitem[Fialkov et al.(2018)]{fialkov2018} Fialkov, A., Barkana, R., \& Cohen, A. 2018, \href{https://arxiv.org/abs/1802.10577}{arXiv:1802.10577}

\bibitem[Furlanetto et al.(2006)]{furlanetto2006} Furlanetto, S. R., Oh, S. P., \& Briggs, F. H., 2006, \href{https://doi.org/10.1016/j.physrep.2006.08.002}{\physrep} 433, 181

\bibitem[G\'orski et al.(2005)]{gorski2005} G\'orski, K.M., Hivon, E., Banday, A.J., et al. 2005, \href{https://doi.org/10.1086/427976}{\apj}, 622, 759

\bibitem[Greig \& Mesinger(2015)]{greig2015} Greig B., \& Mesinger, A. 2015, \href{https://doi.org/10.1093/mnras/stv571}{\mnras}, 449, 4246

\bibitem[Greig \& Mesinger(2017a)]{greig2017a} Greig, B., \& Mesinger, A. 2017, \href{https://doi.org/10.1093/mnras/stw3026}{\mnras}, 465, 4838

\bibitem[Greig \& Mesinger(2017b)]{greig2017b} Greig, B., \& Mesinger, A. 2017, \href{https://doi.org/10.1093/mnras/stx2118}{\mnras}, 472, 2651

\bibitem[Greig et al.(2017)]{greig2017c} Greig, B., Mesinger, A., Haiman, Z., \& Simcoe, R. 2017, \href{https://doi.org/10.1093/mnras/stw3351}{\mnras}, 466, 4239

\bibitem[Guzm\'an et al.(2011)]{guzman2011} Guzm\'an, A. E., May, J., Alvarez, H., Maeda, K. 2011, \href{https://doi.org/10.1051/0004-6361/200913628}{\aaps}, 525, A138

\bibitem[Harker et al.(2012)]{harker2012} Harker, G. J. A., Pritchard, J. R., Burns, J. O., \& Bowman, J. D. 2012, \href{https://doi.org/10.1111/j.1365-2966.2011.19766.x}{\mnras}, 419, 1070

\bibitem[Harker (2015)]{harker2015} Harker, G. J. A. 2015, \href{https://doi.org/10.1093/mnrasl/slv011}{\mnras}, 449, L21

\bibitem[Harker et al.(2016)]{harker2016} Harker, G. J. A., Mirocha, J., Burns, J. O., \& Pritchard, J. R. 2016, \href{https://doi.org/10.1093/mnras/stv2630}{\mnras}, 455, 3829

\bibitem[Haslam et al.(1982)]{haslam1982} Haslam, C. G. T., Salter, C. J., Stoffel, H., \& Wilson, W. E. 1982, \href{http://adsabs.harvard.edu/abs/1982A\%26AS...47....1H}{\aaps}, 47, 1

\bibitem[Kern et al.(2017)]{kern2017} Kern, N. S., Liu, A., Parsons A. R., Mesinger, A., \& Greig, B. 2017, \href{https://doi.org/10.3847/1538-4357/aa8bb4}{\apj}, 848, 1

\bibitem[McGreer et al.(2015)]{mcgreer2015} McGreer, I. D., Mesinger A., D'Odorico V. 2015, \href{https://doi.org/10.1093/mnras/stu2449}{\mnras}, 447, 499

\bibitem[Mesinger \& Furlanetto(2007)]{mesinger2007} Mesinger, A., \& Furlanetto, S. R. 2007, \href{https://doi.org/10.1086/521806}{\apj}, 669, 663

\bibitem[Mesinger et al.(2011)]{mesinger2011} Mesinger, A., Furlanetto, S. R., \& Cen R. 2011, \href{https://doi.org/10.1111/j.1365-2966.2010.17731.x}{\mnras}, 411, 955

\bibitem[Mesinger et al.(2013)]{mesinger2013} Mesinger, A., Ferrara, A., \& Spiegel, D. S. 2013, \href{https://doi.org/10.1093/mnras/stt198}{\mnras}, 431, 621

\bibitem[Mineo et al.(2012)]{mineo2012} Mineo, S., Gilfanov, M., Sunyaev, R. 2012, \href{https://doi.org/10.1111/j.1365-2966.2011.19862.x}{\mnras}, 419, 2095

\bibitem[Mirocha \& Furlanetto(2018)]{mirocha2018b} Mirocha, J. \& Furlanetto, S. R. 2018, \href{https://arxiv.org/abs/1803.03272}{arXiv:1803.03272}

\bibitem[Monsalve et al.(2017a)]{monsalve2017a} Monsalve, R. A., Rogers, A. E. E., Bowman, J. D., \& Mozdzen, T. J. 2017, \href{https://doi.org/10.3847/1538-4357/835/1/49}{\apj}, 835, 49

\bibitem[Monsalve et al.(2017b)]{monsalve2017b} Monsalve, R. A., Rogers, A. E. E., Bowman, J. D., \& Mozdzen, T. J. 2017, \href{https://doi.org/10.3847/1538-4357/aa88d1}{\apj}, 847, 64

\bibitem[Mortlock et al.(2011)]{mortlock2011} Mortlock, D., et al. 2011, \href{https://doi.org/10.1038/nature10159}{\nat}, 474, 616

\bibitem[Mozdzen et al.(2016)]{mozdzen2016} Mozdzen, T. J., Bowman, J. D., Monsalve, R. A., Rogers, A. E. E. 2016, \href{https://doi.org/10.1093/mnras/stv2601}{\mnras}, 455, 3890

\bibitem[Mu\~noz et al.(2015)]{munoz2015} Mu\~noz, J. B., Kovetz, E. D., \& Ali-Ha\"imoud, Y. 2015, \href{https://doi.org/10.1103/PhysRevD.92.083528}{\prd}, 92, 083528

\bibitem[Mu\~noz \& Loeb(2018)]{munoz2018} Mu\~noz, J. B. \& Loeb, A. 2018, \href{https://doi.org/10.1038/s41586-018-0151-x}{\nat}, 557, 684

\bibitem[Planck Collaboration XLVII (2016)]{planck2016} Planck Collaboration XLVII 2016, \href{https://doi.org/10.1051/0004-6361/201628897}{\aap}, 596, A108

\bibitem[Price et al.(2018)]{price2018} Price, D. C. et al. 2018, \href{https://doi.org/10.1093/mnras/sty1244}{\mnras}

\bibitem[Pritchard \& Furlanetto(2007)]{pritchard2007} Pritchard, J. R., \& Furlanetto, S. R. 2007, \href{https://doi.org/10.1111/j.1365-2966.2007.11519.x}{\mnras}, 376, 1680

\bibitem[Pritchard \& Loeb(2010)]{pritchard2010} Pritchard, J. R., \& Loeb, A. 2010, \href{https://doi.org/10.1103/PhysRevD.82.023006}{\prd}, 82, 023006

\bibitem[Rogers \& Bowman(2012)]{rogers2012} Rogers, A. E. E., \& Bowman, J. D. 2012, \href{https://doi.org/10.1029/2011RS004962}{Radio Sci.}, 47, RS0K06

\bibitem[Sathyanarayana Rao et al.(2015)]{sathyanarayana2015} Sathyanarayana Rao, M., Subrahmanyan, R., Udaya Shankar, N., \& Chluba J. \href{https://doi.org/10.1088/0004-637X/810/1/3}{\apj}, 810, 3

\bibitem[Sathyanarayana Rao et al.(2017a)]{sathyanarayana2017a} Sathyanarayana Rao, M., Subrahmanyan, R., Udaya Shankar, N., \& Chluba J. \href{https://doi.org/10.3847/1538-3881/153/1/26}{\apj}, 153, 26

\bibitem[Sathyanarayana Rao et al.(2017b)]{sathyanarayana2017b} Sathyanarayana Rao, M., Subrahmanyan, R., Udaya Shankar, N., \& Chluba J. \href{https://doi.org/10.3847/1538-4357/aa69bd}{\apj}, 840, 33

\bibitem[Singh et al.(2017a)]{singh2017a} Singh, S. et al. 2017, \href{https://doi.org/10.3847/2041-8213/aa831b}{\apjl}, 845, L12

\bibitem[Singh et al.(2017b)]{singh2017b} Singh, S. et al. 2017, \href{https://arxiv.org/abs/1711.11281}{arXiv:1711.11281} 

\bibitem[Sobacchi \& Mesinger(2014)]{sobacchi2014} Sobacchi, E. \& Mesinger, A. 2014, \href{https://doi.org/10.1093/mnras/stu377}{\mnras}, 440, 1662

\bibitem[Switzer \& Liu(2014)]{switzer2014} Switzer, E. R. \& Liu A. 2014 \href{https://doi.org/10.1088/0004-637X/793/2/102}{\apj}, 793, 102

\bibitem[Tashiro et al.(2014)]{tashiro2014} Tashiro, H., Kadota, K., \& Silk, J. 2014, \href{https://doi.org/10.1103/PhysRevD.90.083522}{\prd}, 90, 083522

\bibitem[Tauscher et al.(2018)]{tauscher2018} Tauscher, K., Rapetti, D., Burns, J. O., \& Switzer, E. R. 2018, \href{https://doi.org/10.3847/1538-4357/aaa41f}{\apj}, 853, 187

\bibitem[The Astropy Collaboration et al.(2013)]{astropy2013} The Astropy Collaboration, Robitaille, T. P., Tollerud, E. J., et al. 2013, \href{https://doi.org/10.1051/0004-6361/201322068}{\aap}, 558, A33

\bibitem[Vedantham et al.(2014)]{vedantham2014} Vedantham, H. K., Koopmans, L. V. E., de Bruyn, A. G., et al. 2014, \href{https://doi.org/10.1093/mnras/stt1878}{\mnras}, 437, 1056

\bibitem[Voytek et al.(2014)]{voytek2014} Voytek, T. C., Natarajan, A., J\'auregui Garc\'ia, J. M., Peterson, J. B. \& L\'opez-Cruz, O. 2014, \href{https://doi.org/10.1088/2041-8205/782/1/L9}{\apjl}, 782, L9

\bibitem[Zheng et al.(2017)]{zheng2017} Zheng et al. 2017, \href{https://doi.org/10.1093/mnras/stw2525}{\mnras}, 464, 3486

\end{thebibliography}
\end{document}